\documentclass[twocolumn,prl,showpacs,preprintnumbers,
               superscriptaddress,floatfix]{revtex4}
\usepackage{amsfonts}
\usepackage{graphicx}
\usepackage{graphicx,epsfig,latexsym,amssymb}
\usepackage{multirow,amsmath,array,booktabs}
\usepackage{dcolumn}
\usepackage[section]{placeins}
\usepackage{bm}

\newcommand{\ls}{\left[}
\newcommand{\rs}{\right]}
\newcommand{\pade}{\mbox{Pad$\acute{\text{e}}$}}
\newcommand{\svec}[1]{{\mbox{\boldmath${\rm #1}$}}}

\newcommand{\re}{\nonumber\\}

\renewcommand{\emph}[1]{\textit{\textbf{#1}}}

\begin{document}

\title{Isospin dependence of pseudospin symmetry in nuclear resonant states}
\author{Jian You Guo}
\email{jianyou@ahu.edu.cn}
\affiliation{School of physics and material science, Anhui university, Hefei 230039, P.R.
China}
\author{Xiang Zheng Fang}
\affiliation{School of physics and material science, Anhui university, Hefei 230039, P.R.
China}

\begin{abstract}
The relativistic mean field theory in combination with the
analytic continuation in the coupling constant method is used to
determine the energies and widths of single-particle resonant
states in Sn isotopes. It is shown that there exists clear shell
structure in the resonant levels as appearing in the bound levels.
In particular, the isospin dependence of pseudospin symmetry is
clearly shown in the resonant states, is consistent with that in
the bound states, where the splittings of energies and widths
between pseudospin doublets are found in correlation with the
quantum numbers of single-particle states, as well as the nuclear
mass number. The similar phenomenon also emerges in the spin
partners.
\end{abstract}

\pacs{21.10.Hw, 21.60.Jz, 25.70.Ef, 27.80.+w}\maketitle
 
\section{Introduction}
About 30 years ago a quasi degeneracy was observed in heavy nuclei
between single-nucleon doublets with quantum numbers $(n,l,j=l+1/2)$ and $%
(n-1,l+2,j=l+3/2)$ where $n$, $l$, and $j$ are the radial, the orbital, and
the total angular momentum quantum numbers, respectively\cite%
{Hecht69,Arima69}. The quasi-degenerative states are suggested as the
pseudospin doublets $j=\tilde{l}\pm \tilde{s}$ with the pseudo orbital
angular momentum $\tilde{l}$, and the pseudospin angular momentum $\tilde{s}$%
, and have explained a number of phenomena in nuclear structure
including deformation\cite{Bohr82} and
superdeformation\cite{Dudek87}, magnetic
moment\cite{Trolt94,Stuch992} and identical
bands\cite{Nazar90,Steph908,Zeng91}. Because of these successes,
there have been comprehensive efforts to
understand its origin since the discovery of this symmetry. Blokhin et al.%
\cite{Blokh95} performed a helicity unitary transformation in a
nonrelativistic single-particle Hamiltonian, showed that the transformed
radial wave functions have a different asymptotic behavior, implying that
the helicity transformed mean field acquires a more diffuse surface.
Application of the helicity operator to the nonrelativistic single-particle
wave function maps the normal state ($l$, $s$) onto the pseudo state ($%
\tilde{l}$, $\tilde{s}$), while keeping all other global
symmetries. The same kind of unitary transformation was also
considered earlier by Bahri et al. to discuss the pseudospin
symmetry in the nonrelativistic harmonic oscillator\cite{Bahri92}.
They showed that a particular condition between the coefficients
of spin-orbit and orbit-orbit terms, required to have a pseudospin
symmetry in that nonrelativistic single-particle Hamiltonian, was
consistent with relativistic mean-field (RMF) estimates. Based on
the RMF theories, Ginocchio identifies a possible reason for this;
namely that the symmetry arises from the near equality in
magnitude of an attractive scalar, $S$, and repulsive vector, $V$,
relativistic mean field, $S\sim -V$, in which the nucleons
move\cite{Ginoc979}. They reveal the pseudospin symmetry is exact
when doublets are degenerate. Meng et al.\cite{Meng98,Meng99} show
that pseudospin symmetry is exact when $d\Sigma /dr=0$ where
$\Sigma=V+S$ and the quality of the pseudospin approximation in
real nuclei is connected with the competition between the
pseudo-centrifugal barrier and the
pseudospin-orbital potential. In other works\cite%
{Lalaz98,Ginoc98,Ginoc99,Sugaw98} the pseudospin symmetry has also been
discussed in relation with the arguments and conjectures of Refs.\cite%
{Ginoc979,Meng98,Meng99}. Pseudospin symmetry has also been shown
to be approximately conserved in medium energy nucleon scattering
from even-even nuclei\cite{Ginoc99,Leeb00}. A test of nuclear wave
functions for pseudospin symmetry was done in
Ref.\cite{Ginoc01,Ginoc022}. The structure of radial nodes
occurring in pseudospin levels and a classification for the
intruder levels states which do not have a pseudospin partner in
the limit of pseudospin symmetry was explained as a direct effect
of the behavior of nodes of Dirac bound states\cite{Levia01},
giving support for the relativistic interpretation of nuclear
pseudospin symmetry. Pseudospin symmetry is also investigated for
the relativistic harmonic oscillator, and relativistic Woods-Saxon
and Hulth\'en problem, in which the analytical solutions are
obtained for the bound states of the corresponding Dirac equations
by setting either $\Sigma $ or $\Delta$ $(=V-S)$ to
zero\cite{Chen03,Ginoc04,Lisbo041,Guo02,Guo052,Guo03}. The
pseudospin breaking is shown to be connected with the mean-field
shaped by the parameters of the harmonic oscillator
potential\cite{Guo051}, or the Woods-Saxon
potential\cite{Alber012}. All these indicate the pseudospin
symmetry is a good approximate for the single-particle bound
states of atomic nuclei. However for the resonant states, which
are thought to play a important role in formation of the exotic
phenomena such as halo and skin\cite{Meng9698,Sandu00,Cao02}, the
symmetry has not been examined in detail for real nuclei. In
recent years, many techniques have been developed to study
resonant states, such as the R-matrix theory\cite{Wigner47}, the
extended R-matrix theory\cite{Hale87}, the K-matrix
theory\cite{Humblet91}, S-matrix theory\cite{Taylor72}, the
real stabilization method\cite{Hazi70}, the complex scaling method\cite{Ho83}%
, and so on. Compared with these methods, the analytic
continuation in the coupling constant (ACCC) method proposed by
Kukulin\cite{Kukulin89} is more effective and numerically quite
simple, has been applied to investigate the energies and widths
for resonant states in the framework of the
non-relativistic\cite{Tanaka99,Aoyama02,Cattapan00} and relativistic theory%
\cite{Yang01,Zhang05}. Based on the Schr\"odinger and Dirac
equations, we have investigated the stability and convergence of
the energies and widths for single-particle resonant states with
square-well, harmonic oscillator and Woods-Saxon potentials, and
discussed their dependence on the coupling constant interval and
the Pad\'e approximant (PA) order\cite{Zhang03,Zhang04}.
Particularly, we have studied the pseudospin symmetry in the
resonant states for Dirac-Woods-Saxon problem, and shown the
correlations of pseudospin splittings with the shapes of
Woods-Saxon mean field\cite{Guo053}. Considering that the RMF
theory is a more microscopic one, and has gained considerable
success in describing many nuclear phenomena for the stable nuclei
as well as nuclei even far from stability, here we intend to study
the resonant states including their pseudospin and spin symmetries
by the ACCC method in the framework of RMF theory, as done for the
bound states in the RMF theory\cite{Meng99}.

\section{Theoretical Framework}
The basic ansatz of the RMF theory is a Lagrangian density whereby
nucleons are described as Dirac particles which interact via the
exchange of various mesons (the scalar $\sigma$, vector $\omega$
and iso-vector vector $\rho$) and the photon~\cite{Meng982}
\begin{eqnarray}
\begin{array}{cc}
{\cal L} &= \bar \psi (i\rlap{/}\partial -M) \psi +
        \,{1\over2}\partial_\mu\sigma\partial^\mu\sigma-U(\sigma)
        -{1\over4}\Omega_{\mu\nu}\Omega^{\mu\nu} \\
        \ &+ {1\over2}m_\omega^2\omega_\mu\omega^\mu
        -{1\over4}{\svec R}_{\mu\nu}{\svec R}^{\mu\nu} +
        {1\over2}m_{\rho}^{2} \svec\rho_\mu\svec\rho^\mu
        -{1\over4}F_{\mu\nu}F^{\mu\nu} \\
        & -  g_{\sigma}\bar\psi \sigma \psi~
        -~g_{\omega}\bar\psi \rlap{/}\omega \psi~
        -~g_{\rho}  \bar\psi
        \rlap{/}\svec\rho
        \svec\tau \psi
        -~e \bar\psi \rlap{/}\svec A \psi,
\label{Lagrangian}
\end{array}
\end{eqnarray}
where $M$ is the nucleon mass and $m_\sigma$ ($g_\sigma$),
$m_\omega$ ($g_\omega$), and $m_\rho$ ($g_\rho$) are the masses
(coupling constants) of the respective mesons. A nonlinear scalar
self-interaction $U(\sigma)~=~\dfrac{1}{2} m^2_\sigma \sigma^2
~+~\dfrac{g_2}{3}\sigma^3~+~\dfrac{g_3}{4}\sigma^4$ of the
$\sigma$ meson has been included~\cite{BB77}. The field tensors
for the vector mesons are given as
\begin{eqnarray}
\left\{
\begin{array}{lll}
   \Omega^{\mu\nu}   &=& \partial^\mu\omega^\nu-\partial^\nu\omega^\mu, \\
   {\svec R}^{\mu\nu} &=& \partial^\mu{\svec \rho}^\nu
                        -\partial^\nu{\svec \rho}^\mu
                        - g^{\rho} ( {\svec \rho}^\mu
                           \times {\svec \rho}^\nu ), \\
   F^{\mu\nu}        &=& \partial^\mu \svec A^\nu-\partial^\nu \svec A^\mu.
\end{array}   \right.
\label{tensors}
\end{eqnarray}

The classical variation principle gives the following equations of
motion
\begin{equation}
   [ {\svec \alpha} \cdot {\svec p} +
     V_V ( {\svec r} ) + \beta ( M + V_S ( {\svec r} ) ) ]
     \psi ~=~ \varepsilon\psi,
\label{spinor1}
\end{equation}
for the nucleon spinors, where $\varepsilon$ and $\psi$ are the
single-particle energy and spinor wave function respectively, and
the Klein-Gordon equations
\begin{eqnarray}
\left\{
\begin{array}{lll}
  \left( -\Delta \sigma ~+~U'(\sigma) \right ) &=& g_\sigma\rho_s,
\\
   \left( -\Delta~+~m_\omega^2\right )\omega^{\mu} &=&
                    g_\omega j^{\mu} ( {\svec r} ),
\\
   \left( -\Delta~+~m_\rho^2\right) {\svec \rho}^{\mu}&=&
                    g_\rho \svec j^{\mu}( {\svec r} ),
\\
          -\Delta~ A_0^{\mu} ( {\svec r} ) ~ &=&
                           e j_{\rho}^{\mu}( {\svec r} ),
\end{array}  \right.
\label{mesonmotion}
\end{eqnarray}
for the mesons, where
\begin{eqnarray}
\left\{
\begin{array}{lll}
   V_V( {\svec r} ) &=&
      g_\omega\rlap{/}\omega + g_\rho\rlap{/}\svec\rho\svec\tau
         + \dfrac{1}{2}e(1-\tau_3)\rlap{\,/}\svec A , \\
   V_S( {\svec r} ) &=&
      g_\sigma \sigma , \\
\end{array}
\right.
\label{vaspot}
\end{eqnarray}
are the vector and scalar potentials respectively and the source
terms for the mesons are
\begin{eqnarray}
\left\{
\begin{array}{lll}
   \rho_s &=& \sum\limits_{i=1}^A \bar\psi_i \psi_i,
\\
   j^{\mu} ( {\svec r} ) &=&
               \sum\limits_{i=1}^A \bar \psi_i \gamma^{\mu} \psi_i,
\\
   \svec j^{\mu}( {\svec r} ) &=&
          \sum\limits_{i=1}^A \bar \psi_i \gamma^{\mu} \svec \tau
          \psi_i,
\\
   j^{\mu}_p ( {\svec r} ) &=&
      \sum\limits_{i=1}^A \bar \psi_i \gamma^{\mu} \dfrac {1 - \tau_3} 2
      \psi_i .
\end{array}
\right.
\label{mesonsource}
\end{eqnarray}
It should be noted that the contribution of negative energy states
are neglected, i.e., the vacuum is not polarized. Moreover, the
mean field approximation is carried out via replacing meson field
operators in Eq. (\ref{spinor1}) by their expectation values,
since the coupled equations Eq. (\ref{spinor1}) and Eq.
(\ref{mesonmotion}) are nonlinear quantum field equations and
their exact solutions are very complicated. In this way, the
nucleons are assumed to move independently in the classical meson
fields. The coupled equations can be solved self-consistently by
iteration.

For spherical nuclei, the potential of the nucleon and the sources
of meson fields depend only on the radial coordinate $r$. The
spinor is characterized by the angular momentum quantum numbers
$l$, $j$, $m$, the isospin $t = \pm \dfrac 1 2$ for neutron and
proton respectively. The Dirac spinor has the form
\begin{equation}
   \psi ( \svec r ) =
      \left( { {\mbox{i}  \dfrac {G_{lj}(r)} r {Y^l _{jm} (\theta,\phi)} }
      \atop
       { \dfrac {F_{lj}(r)} r (\svec\sigma \cdot \hat {\svec r} )
       {Y^l _{jm} (\theta,\phi)} } }
      \right) \chi _{t}(t),
\label{reppsi}
\end{equation}
where $Y^l _{jm} (\theta,\phi)$ are the spinor spherical
harmonics. The radial equation of the spinor, i.e. Eq.
(\ref{spinor1}), can be reduced to ~\cite{Meng982}
\begin{eqnarray}
  \label{Dirac-r}
         (-\dfrac{\partial}{\partial
          r}+\dfrac{\kappa}{r})F_{lj}(r)+\ls M+V_p(r)\rs
          G_{lj}(r)&=&\varepsilon G_{lj}(r),\re
         (\dfrac{\partial}{\partial r}+\dfrac{\kappa}{r})G_{lj}(r)-\ls
         M-V_m(r)\rs F_{lj}(r)&=&\varepsilon F_{lj}(r),
\end{eqnarray}
in which $V_p(r)=V_V(r)+V_S(r)$, $V_m(r)=V_V(r)-V_S(r)$, and
$\kappa=(-1)^{j+l+1/2}(j+1/2)$. The meson field equations can be
reduced to
\begin{equation}
    \left( \frac {\partial^2} {\partial r^2}  - \frac 2 r \frac
         {\partial}   {\partial r} + m_{\phi}^2 \right)\phi = s_{\phi} (r),
    \label{Ramesonmotion}
\end{equation}
where $\phi = \sigma$, $\omega$, $\rho$, and photon ( $m_{\phi} =
0$ for photon). The source terms read
\begin{eqnarray}
  s_{\phi} (r) = \left\{
    \begin{array}{ll}
      -g_\sigma\rho_s - g_2 \sigma^2(r)  - g_3 \sigma^3(r)
           & { \rm for ~ the ~  \sigma~  field }, \\
      g_\omega \rho_v    & {\rm for ~ the ~ \omega ~ field}, \\
      g_{\rho}  \rho_3(r)       & {\rm for~ the~ \rho~ field},\\
      e \rho_c(r)  & {\rm for~ the~ Coulomb~ field}, \\
   \end{array} \right.
\end{eqnarray}
and
\begin{eqnarray}
\left\{
\begin{array}{lll}
   4\pi r^2 \rho_s (r) &=& \sum\limits_{i=1}^A ( |G_i(r)|^2 - |F_i(r)|^2 ), \\
   4\pi r^2 \rho_v (r) &=& \sum\limits_{i=1}^A ( |G_i(r)|^2 + |F_i(r)|^2 ), \\
   4\pi r^2 \rho_3 (r) &=& \sum\limits_{p=1}^Z ( |G_p(r)|^2 + |F_p(r)|^2
   )\\&&-\sum\limits_{n=1}^N ( |G_n(r)|^2 + |F_n(r)|^2 ), \\
   4\pi r^2 \rho_c (r) &=& \sum\limits_{p=1}^Z ( |G_p(r)|^2 + |F_p(r)|^2 ) . \\
\end{array}
\right. \label{mesonsourceS}
\end{eqnarray}

By solving Eqs.~(\ref{Dirac-r}) and (\ref{Ramesonmotion}) in a
meshed box of size $R_0$ self-consistently, one can calculate the
ground state properties of a nucleus. The vector potential
$V_V(r)$ and the scalar potential $V_S(r)$, energies and wave
functions for bound states are also obtained. By increasing the
attractive potential as $V_p(r)\rightarrow\lambda V_p(r)$, a
resonant state will be lowered and becomes a bound state if the
coupling constant $\lambda$ is large enough. Near the branch point
$\lambda_0$, defined by the scattering threshold
$k(\lambda_0)=0$~\cite{Kukulin89}, the wave number $k(\lambda)$
behaves as
\begin{eqnarray}
   k(\lambda) \sim \left\{
   \begin{array}{l@{\mbox{~~~}}l}
         i\sqrt{\lambda-\lambda_0},        &  l>0 ,\\
         i(\lambda-\lambda_0),            &  l=0.
   \end{array}\right.
\end{eqnarray}
These properties suggest an analytic continuation of the wave
number $k$ in the complex $\lambda$ plane from the bound-state
region into the resonance region by $\pade$ approximant of the
second kind (PAII)~\cite{Kukulin89}
\begin{equation}
    \label{pade-e}
    k(x)\approx k^{[L,N]}(x)= i \dfrac{c_0+c_1 x+c_2x^2+\ldots +c_Lx^L}{ 1+d_1 x+d_2 x^2+\ldots+d_N x^N},
\end{equation}
where $x\equiv\sqrt{\lambda-\lambda_0}$, and $c_0, c_1,\ldots,
c_L, d_1, d_2,\ldots,d_N$ are the coefficients of PA. These
coefficients can be determined by a set of reference points $x_i$
and $k(x_i)$ obtained from the Dirac equation with
$\lambda_i>\lambda_0,~i=1,2,...,L+N+1$. With the complex wave
number $k(\lambda=1)= k_r + i k_i$, the resonance energy $E$ and
the width $\Gamma$ can be extracted from the relation $
\varepsilon=E-i \dfrac{\Gamma}{2} ~~ (E,\Gamma\in \mathbb{R})$ and
$k^2=\varepsilon^2-M^2$, i.e.,
\begin{eqnarray}
    \label{E-W}
    E      &=& \sqrt{\dfrac{\sqrt{(M^2+k_r^2-k_i^2)^2+4k_r^2k_i^2}+(M^2+k_r^2-k_i^2)}{2}}- M , \re
    \Gamma &=& \sqrt{2\sqrt{(M^2+k_r^2-k_i^2)^2+4k_r^2k_i^2}-2(M^2+k_r^2-k_i^2)}.
\end{eqnarray}
In the non-relativistic limit ($k\ll M$), Eq. (\ref{E-W}) reduces
to
\begin{equation}
    E=\dfrac{k_r^2-k_i^2}{2M},~~~~~~~~~~~~~~~~~
    \Gamma=\dfrac{2k_rk_i}{M}.
\end{equation}
\begin{figure}[th]
\centering \includegraphics[width=8.5cm]{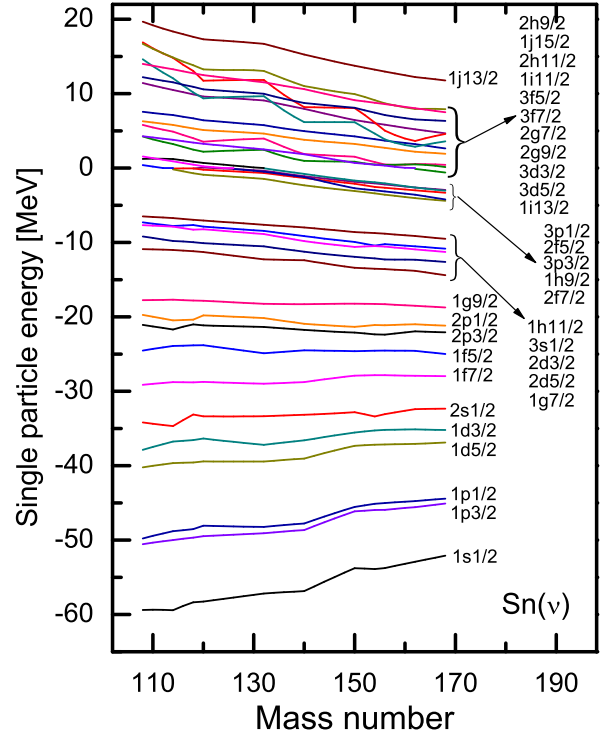}
\vspace{-10pt} \caption{ The single-particle energies of neutron
as the function of mass number for Sn isotopes, where the energies
of resonant states obtained by the ACCC method in the RMF theory
with the interactions NL3. } \label{levels}
\end{figure}
\begin{figure}[th]
\centering \includegraphics[width=8.5cm]{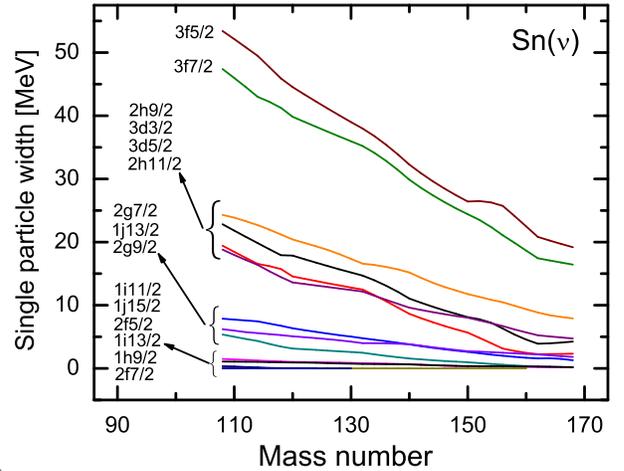}
\vspace{-5pt} \caption{The same as Fig.\ref{levels}, but for the
single-particle widths of neutron.}
\end{figure}

\section{Numerical details and results}
With the technique represented above, we first single out the
resonant states in the continuum for Sn isotopes, and then analyze
their properties including the pseudospin and spin symmetries.
Following Refs.\cite{Zhang03,Zhang04,Zhang05,Guo053}, the radial
Dirac equations (\ref{Dirac-r}) are solved in a standard way by
the shooting method with a step size of 0.05 fm using proper
boundary conditions in a spherical box of radius $R=20$
fm\cite{Meng982}. The branch point $\lambda_0$ is
determined in a self-consistent way as in Ref.\cite{Kukulin89}. In correspondence to $%
L+N+1$ different coupling constant values $\lambda_i>\lambda_0$,
we solve the bound state problem for $\lambda_i\Sigma(r)$, and
obtain the eigenvalues $\varepsilon_i$ from the Dirac equations.
The solutions of the linear equations are the coefficients in the
PA expression (\ref{pade-e}). With the coefficients in
Eq.(\ref{pade-e}), the wave number as a function of the coupling
constant $k(\lambda)$ can be obtained. Then the energy and width
of a resonant state can be extracted from the complex wave number
$k(\lambda=1)$. Considering that there are a little influences of
the coupling constant interval and the Pad\'e approximant order on
the calculated results in the ACCC method as shown in
Ref.\cite{Guo053}, we first check the stability and convergence of
this method in the present framework to choose reasonably the
coupling constant interval and the Pad\'e approximant order for Sn
isotopes. Similar to the Figs.2 and 3 in Ref.\cite{Guo053}, the
energies and widths of resonant states are found to be
considerable stable in a fairly large coupling constant interval
for fixed order, especially for the calculations with (5,5) PAs.
Based on the fact, all the energies and widths for resonant states
in the following are taken from the calculations of (5,5) PA with
$\lambda_b-\lambda_0=1$ as done in Ref.\cite{Guo053}, in which the
branch point $\lambda_0$ and $\lambda_b$ are respectively the left
and right edge of the coupling constant interval in the Pad\'e
approximant.
\begin{figure}[th]
\centering \includegraphics[width=8.5cm]{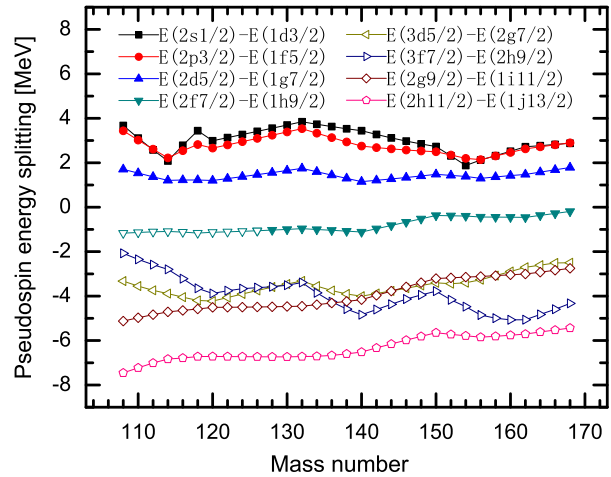}
\vspace{-10pt} \caption{The energy splittings of pseudospin
doublets as the function of mass number for Sn isotopes, where the
filled and opened marks represent respectively for the cases of
bound and resonant states. }
\end{figure}
\begin{figure}[th]
\centering \includegraphics[width=8.5cm]{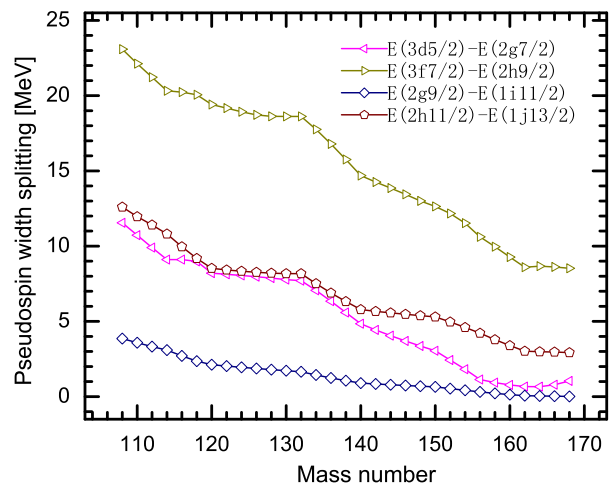}
\vspace{-10pt} \caption{The same as Fig.3, but for the width
splittings of pseudospin doublets for Sn isotopes.}
\end{figure}

To see the behavior of the single-particle resonant states and
their isospin dependence as well as comparison with the case of
bound states, the energy as a function of the mass number are
displayed in Fig.1 for the Sn isotopes. Except for several small
kinks, the energies are shown to decrease monotonously with the
increasing isospin for all the resonant states, which is in
agreement with that for the bound states near the threshold,
different from that for the bound states far from the threshold
where the change is altered with a contrary trend as the level
$1s_{1/2}$. For the resonant states $3d_{3/2}$,$3d_{5/2}$,
$3f_{5/2}$, and $3f_{7/2}$, the energies decrease more quickly,
and happen to cross with their neighbor at several Sn nuclei,
which does not appear in the bound states. Particularly, there are
several kinks in these levels as appearing in the bound levels at
a less extent, e.g.: the $1s_{1/2}$,$2s_{1/2}$, $2p_{1/2}$, and
$2p_{3/2}$. These kinks always appear at several special positions
for whether the bound or resonant states, may reflect some special
structure existing in these nuclei. Besides of the kinks, several
large gaps are found in the resonant levels as appearing in the
bound levels. The large energy gap between the levels $2f_{7/2}$
and $1h_{11/2}$ remain hardly change whether these levels are
unbound or lowered to bound. The same phenomenon also appear in
the resonant levels between $1j_{13/2}$ and $2h_{9/2}$. From these
phenomena we see that the resonant levels also provide the
detailed information on nuclear shell structure.

Compared with the energies, the widths of the resonant states also
decrease monotonously with the increasing isospin, which can be
seen in Fig.2. Especially for the resonant states $3d_{3/2}$,
$3d_{5/2}$, $3f_{5/2}$, and $3f_{7/2}$, the behavior of decreasing
more quickly, even crossing with their neighbor in the energies
also appears in the widths. Moreover, the shell structure is
clearly shown with several large gaps in the widths for several
different resonant states. The large energy gap between $2h_{9/2}$
and $3f_{7/2}$ decreases with the increasing isospin. The same
phenomenon also appears between the levels  $1i_{11/2}$ and
$2g_{9/2}$. In comparison with the energies, the widths decrease
in general more quickly, especially for the $3f_{5/2}$ and
$3f_{7/2}$.

With these knowledge on the energies and widths, we analyze the
symmetry existing in these resonant states. As identified to be a
good approximation in the bound states, the pseudospin symmetry
worths being examined for the single-particle resonant states. In
Ref.\cite{Guo053}, the symmetry in the resonant states has been
investigated by solving the Dirac equation with Woods-Saxon
potentials in combination with the ACCC method, where the
pseudospin breaking is shown in correlation with the shapes of
nuclear mean-field. In order to examine the symmetry in the
framework of RMF theory, and study their isospin dependence, which
has been discussed in the bound states\cite{Meng99}, we analyze
the relation between the pseudospin splitting and the nuclear mass
number for the resonant states. The dependence is shown in Fig.3
for the Sn isotopes, where the energy splittings of bound states
are also exhibited for comparison. From Fig.3, it can be seen that
the pseudospin symmetry in the resonant states is correlated with
the quantum numbers of single-particle states, and the pseudospin
splittings are different from the different pseudospin partners,
which is in agreement with the case of bound states\cite{Alber012}
as well as the case of resonant states in Dirac Woods-Saxon mean
field approximation\cite{Guo053}. As far as the isospin dependence
of the pseudospin orbital splitting is concerned, the splittings
between the resonant doublets give a monotonous decreasing
behavior with the increasing isospin with only the $3f_{7/3}$ and
$2h_{9/2}$ partner exception. Particularly for the $2g_{9/2}$ and
$1i_{11/2}$ partner, the pseudospin splitting in $^{168}$Sn is
only half of that in $^{108}$Sn. Just as we expected, the
pseudospin symmetry in neutron-rich nuclei is better, which agrees
the case of the bound levels. Besides of the isospin dependence,
the pseudospin splitting is also correlated with the radial,
orbital, and total angler mentum quantum numbers. Such as the
splitting between the $2h_{11/2}$ and $1j_{13/2}$ partner is 2 MeV
larger than that between the $2g_{9/2}$ and $1i_{11/2}$ partner.
In comparison with the resonant states, the situation in the bound
pseudospin partners is more complicated, but the pattern is more
or less the same, i.e., a decreasing behavior with the increasing
isospin with some exceptions. For example, the splittings between
the levels $2s_{1/2}$ and $1d_{3/2}$, and between the levels
$2p_{3/2}$ and $1f_{5/2}$ show a decreasing trend in a complicated
way, while that $2d_{5/2}$ and $1g_{7/2}$ presents a minor
variation in the whole Sn isotope chain. It should be mentioned
for the $2f_{7/2}$ and $1h_{9/2}$ partner, the same trend for the
change of pseudospin splitting is seen whether they are bound or
evolved to unbound states. For all the pseudospin partners with
the same radial quantum number, the energy splitting is discovered
to evolve from $E_{n,l,j=l+1/2}>E_{n-1,l+2,j=l+3/2}$ to
$E_{n,l,j=l+1/2}<E_{n-1,l+2,j=l+3/2}$ with the increasing orbital
angular momentum. This inversion of pseudospin splittings is
observed experimentally, was found in
Refs.\cite{Meng99,Marco00,Marco01,Lalaz98} in the bound states,
also appears in the resonant states here.

For width (to see Fig.4), similar to the energy, the pseudospin
splittings decrease monotonously with the increasing mass number
with no exception. The widths for all the single particle states
with quantum numbers ($n,l,j=l+1/2$) are systemically larger than
that of their pseudospin partners ( $n-1,l+2,j=l+3/2$), which
implies that the resonant states with higher orbital angular
momentum hold longer decay time than their pseudospin partners. It
is because that the orbital angular momentum is smaller, the
centrifugal barrier is lower, the resonant width is larger. Hence,
even the energy is fully degenerate for the pseudospin doublet,
their decay life maybe still different, which is noticeable to
explore the resonant state from experiment. From these studies, we
see that the pseudospin symmetry remains a good approximation for
both stable and exotic nuclei. A better pseudospin symmetry can be
expected for the orbital near the threshold, particularly for
nuclei near the particle drip line.
\begin{figure}[th]
\centering \includegraphics[width=8.5cm]{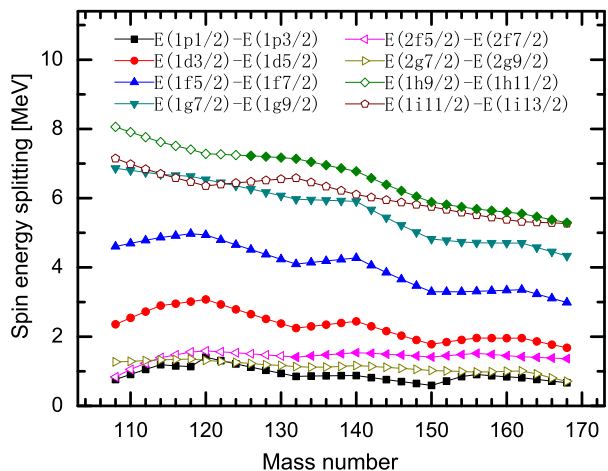}
\vspace{-10pt} \caption{The energy splittings of spin doublets as
the function of mass number for Sn isotopes, where the filled and
opened marks represent respectively to the cases of bound and
resonant states.}
\end{figure}
\begin{figure}[th]
\centering \includegraphics[width=8.5cm]{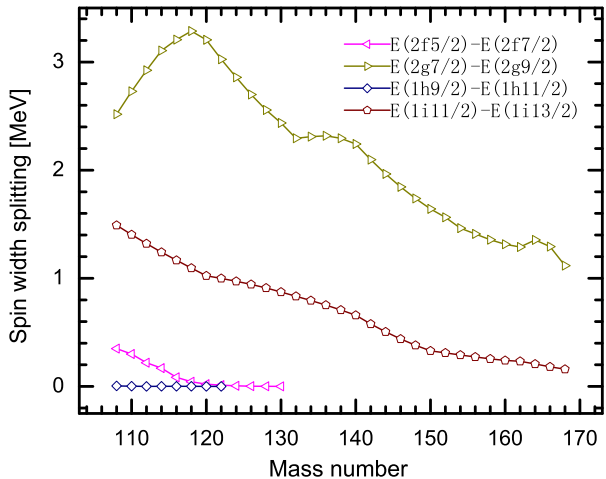}
\vspace{-10pt} \caption{The same as Fig.5, but for the width
splittings of spin doublets for Sn isotopes.}
\end{figure}

In order to compare with the pseudospin symmetry, the spin orbital
splittings as a function of the mass number are plotted in Fig.5
for the Sn isotopes, where a more distinct decreasing behavior
with the increasing mass number is clearly seen for both the bound
and unbound spin partners with few exceptions. It shows that the
spin symmetry in neutron-rich nuclei is also better such as the
pseudospin symmetry. In addition, the spin orbital splitting is
also correlated with the quantum numbers of single-particle
states. For the spin doublets with the same radial quantum number,
the energy splitting increases with the increasing orbital angular
momentum for the bound states, which is contrary to the resonant
states. Compared with pseudospin symmetry, the inversion of spin
splitting has not been found for whether the bound or unbound spin
partners. The energies for the single particle states with quantum
numbers ($n,l,j=l-1/2$) are systemically larger than that of their
spin partners ($n,l,j=l+1/2$), which indicates that the
single-particle states with higher total angular momentum are more
stable than their spin partners. It agrees the experimental facts
as observed in the nuclear bound states. For the width, similar to
the pseudospin symmetry, a monotonous decreasing behavior with the
increasing isospin is seen in Fig.6 with a bit special in several
nuclei. The widths for the single particle states with quantum
numbers ($n,l,j=l-1/2$) are systemically larger than that of their
spin partners ( $n,l,j=l+1/2$), which implies that the resonant
states with higher total angular momentum hold longer decay time
than their spin partners, is similar to the case of pseudospin
symmetry. From these results we see that the spin orbital
splitting is also very sensitive to the quantum numbers of
single-particle states. The spin symmetry remains a better
approximation for the bound states with lower orbital angular
momentum, while for resonant states with higher orbital angular
momentum, particularly for nuclei near the particle drip line.

\section{Conclusion}
The relativistic mean field theory in combination with the
analytic continuation in the coupling constant method is adopted
to determine the energies and widths of single-particle resonant
states for the Sn isotopes, where the energies are found to
decrease monotonously with the increasing isospin with several
small kinks, is in agreement with the case of bound states near
the threshold. In addition, the shell structure is seen clearly in
the resonant levels with several large gaps as appearing in the
bound levels. In particular, the pseudospin and spin symmetries in
the resonant states are investigated and compared with that in the
bound states, where the symmetries are found to be correlated with
the quantum numbers of single-particle states, which is similar to
the case of bound states. As far as the isospin dependence is
concerned, the pseudospin and spin splittings between the resonant
doublets give a monotonous decreasing behavior with the increasing
isospin with few exceptions. The same phenomenon also appear
between the bound doublets. The trend of energy splittings with
the orbital angular momentum is found to be consistent with the
case of bound for all the pseudospin partners, which is not
consistent with the spin partners. The inversion of
energy-splitting appears in several pseudospin partners of
resonant states, which agrees with the experimental observation,
as well as the case of bound states and the nonrelativistic
prediction, while the inversion does not appear in the spin
doublets. The tendency of width splittings with the parameters is
displayed to be agreeable for all the pseudospin partners, as well
as the spin partners. The widths for the single particle states
with higher orbital (total) angular momentum are systemically
larger than that of their pseudospin (spin) partners, which show
that even the fully degenerate pseudospin (spin) doublets may
exist different decay time, making them worthy of experimental
attention.

\begin{acknowledgments}
This work was partly supported by the National Natural Science
Foundation of China under Grant No. 10475001, the Program for New
Century Excellent Talents in University of China, and the
Excellent Talents Foundation in University of Anhui Province in
China.
\end{acknowledgments}

\end{document}